\documentclass[11pt,a4paper]{article}

\usepackage[utf8]{inputenc}
\usepackage{amsmath}
\usepackage[mathcal]{euscript}
\usepackage{latexsym}
\usepackage{textcomp}
\usepackage{slashed}
\usepackage{tikz}
\usepackage{xspace}
\usepackage{setspace}
\usepackage[a4paper]{geometry}
\usepackage[numbers,sort&compress]{natbib}
\newcommand{\beq}{\begin{equation}}
\newcommand{\eeq}{\end{equation}}
\newcommand{\bqa}{\begin{eqnarray}}
\newcommand{\eqa}{\end{eqnarray}}
\newcommand{\bite}{\begin{itemize}}
\newcommand{\eite}{\end{itemize}}
\newcommand{\bd}{\begin{displaymath}}
\newcommand{\ed}{\end{displaymath}}
\newcommand{\bcen}{\begin{center}}
\newcommand{\ecen}{\end{center}}

\def\d#1{D_{#1}}

\def\Formcalc{{{\sc FormCalc}}}

\def\Gosam{{{\sc GoSam}}}
\def\GOSAM{{{\sc GoSam}}}
\def\gosam{{{\sc GoSam}}}
\def\gosamtwo{{{\sc GoSam 2.0}}}

\def\form{{{\sc form}}}

\def\openloops{{{\sc OpenLoops}}}
\def\madloop{{{\sc MadLoop}}}

\def\samurai{{{\sc samurai}}}
\def\SAMURAI{{{\sc samurai}}}
\def\Sherpa{{{\sc Sherpa}}}

\def\cuttools{{{\sc CutTools}}}
\def\C++{{{\sc c++}}}

\def\Golem{{{\sc Golem95C}}}

\def\Ninja{{{\sc Ninja}}}
\def\MadLoop{{{\sc MadLoop}}}
\def\Madspin{{{\sc MadSpin}}}
\def\Pythia{{{\sc Pythia}}}
\def\PB{{{\sc Powheg Box}}}

\def\amcnlomglong{{{\sc MadGraph5\_aMC@NLO}}}
\def\amcnlo{{{\sc MG5\_aMC}}}
\def\amc{{{\sc MG5\_aMC}}}

\def\mgamc{{{\sc MG5\_aMC}}}

\def\ttH{$t\bar{t}H$}
\def\ttyy{$t\bar{t}\gamma\gamma$}

\begin{document}

\begin{center}
%\begin{spacing}{1.5}

{\Large\bf Integrand-reduction techniques for NLO and beyond} \\
%\end{spacing}
\vskip 15mm

{\bf Giovanni Ossola}\\
\small{ \tt gossola@citytech.cuny.edu}
\\[1em]
{\small {\it  New York City College of Technology, 
  City University of New York, \\ 300 Jay Street, Brooklyn NY 11201, USA} \\

  \vspace{0.2cm}

{\it The Graduate School and University Center, 
  City University of New York,  \\ 
  365 Fifth Avenue, New York, NY 10016, USA}
}
\end{center}

\vspace{0.9cm}

\begin{abstract}
\noindent  After a brief general introduction about the integrand-reduction method, we will review the main 
features of the \gosamtwo\ automated framework for one-loop calculations and illustrate its application
to SM processes involving the production of massive particles in
conjunction with jets and photons. These results have been obtained 
by interfacing \gosam\ with different Monte Carlo frameworks, thus 
combining the NLO calculation with parton shower effects.
In the second part of the presentation, we will focus on the applications of \gosam\ beyond NLO.
The code has already been used within NNLO calculations for the computation of real-virtual contributions and for the evaluation of the hard functions needed by approximate NNLO and resummation techniques. 
We will finally briefly discuss a promising approach for the reduction of scattering amplitudes beyond one loop based on integrand
reduction via multivariate polynomial division.
\end{abstract}

\vspace{1.5cm}

\begin{center}
%\begin{spacing}{1.5}
{\it to appear in the proceedings of the} \\

\vspace{0.2cm}

12th International Symposium on Radiative Corrections (Radcor 2015) \\ {\small \it and} \\ LoopFest XIV (Radiative Corrections for the LHC and Future Colliders)\\

\vspace{0.2cm}

                 15-19 June  2015\\
                 UCLA Department of Physics \& Astronomy \\ Los Angeles, CA, USA
%\end{spacing}
\end{center}

%\vspace{0.3cm}
\newpage

\section{Introduction}

In order to obtain precise phenomenological prediction from particle theory, much needed by collider experiments to confirm the current understanding of subatomic physics and possibly to shed light on potential discrepancies between the experimental results and the theoretical models, we need efficient techniques and algorithms to compute scattering amplitudes for a wide variety of processes. 
Scattering amplitudes can be studied by analyzing their symmetries and analytic properties, and a better understanding of their mathematical structure naturally provides the theoretical framework to develop new approaches for their evaluation, and ultimately more efficient algorithms to compute physical cross sections and differential distributions.  

In this interplay between theoretical prediction and experimental data, the precision of such calculations should match the precision of the measurements. Since leading-order (LO) results are affected by large uncertainties, theoretical predictions are not reliable without including the contribution of higher orders. For several analyses, even next-to-leading-order (NLO) accuracy is not sufficient, thus forcing the theorist to tackle extremely challenging next-to-next-to-leading-order (NNLO) calculations,  resort to resummation techniques to properly include all the effects of specific kinematic configurations, or to study clever way to obtain approximate NNLO results, which incorporate most of the physical effects, without the need of a full NNLO computation.

The scope of this talk is to summarize the progress in the evaluation of scattering amplitudes obtained by means of integrand-level techniques, in particular the OPP reduction algorithm, the $d-$dimensional decomposition of scattering amplitudes, and the integrand reduction via multivariate polynomial division. We will start by reviewing the main features of the {\gosam} framework for the automated computation of one-loop amplitudes and comment on some of the recent NLO results obtained using it.  Since {\gosam} generates and evaluates only the virtual part of NLO amplitudes, it is mandatory to interface it with Monte Carlo tools to produce physical results, such as cross sections and differential distributions. We will show examples of applications with particular focus on some recent results obtained by interfacing {\gosam} and \amcnlo.

In the second part of this presentation, we will describe the applications of \gosam\ beyond NLO calculation. In particular, 
\gosam\  has already been used  for the computation of real-virtual contributions within NNLO calculations and for the evaluation of the hard functions needed by approximate NNLO and resummation techniques.  As a forthcoming development, which could be relevant for NNLO automation within an improved version of the code, we will briefly discuss the evaluation of scattering amplitudes beyond one loop based on integrand reduction via multivariate polynomial division.

\section{Integrand Reduction at One Loop}
\label{integrand}

The evaluation of the one-loop diagrams can be performed by decomposing each Feynman integral in terms of a finite set of scalar master integrals~\cite{Passarino:1978jh} plus an additional rational function, known in the literature as \emph{rational part}, which depends on the masses and momenta of the specific process. 
%The calculation of virtual amplitudes can be visualized in terms of three tasks:
%i) the \emph{generation} of the unintegrated amplitudes ${\cal A}$, namely their numerator functions  ${\cal N}(q)$ and the list of denominators $\db{i}$;
%ii) the \emph{reduction} of the amplitude to determine all coefficients multiplying each of the MIs and the rational term ${\cal R}$; 
%iii) the \emph{evaluation of the MIs} which, multiplied by the coefficients obtained in the reduction, provide the final result for the amplitudes. 
Since all scalar integrals are known and readily available in public codes~\cite{vanOldenborgh:1990yc}, the main problem in the evaluation of scattering amplitudes resides in the stable and efficient extraction of all the coefficients which multiply each master integral. 
 
During the past decade, a successful approach to one-loop calculation was developed by merging the idea of four-dimensional unitarity-cuts~\cite{Bern:1994zx,Britto:2004nc} with the understanding of the universal algebraic form of any one-loop scattering amplitudes, as  provided by the OPP method~\cite{delAguila:2004nf, Ossola:2006us, Ossola:2007bb, Ossola:2008xq}.  
In this approach, the coefficients in front of the one-loop MIs can be determined by solving a system of algebraic equations that are obtained by the numerical evaluation of the unintegrated numerator functions at explicit values of the loop-variable and the knowledge of the most general polynomial structure of the integrand itself. 

Such systems of equations become particularly simple when all expressions are evaluated at the complex values of the integration momentum for which a given set of inverse propagators vanish, that define the so-called quadruple, triple, double, and single cuts. This provides a strong connection between the OPP method, and integrand-reduction techniques in general, and generalized unitarity methods, where the on-shell conditions are imposed at the integral level.

\paragraph{Integrand-level Reduction in Four Dimensions} 
The algorithm was originally developed in four dimensions~\cite{Ossola:2006us, Ossola:2007bb, Ossola:2008xq}. 
%According to this approach, the numerator function $N(q)$  which appears in the integrand for any one-loop scattering amplitudes has a universal mathematical structure, independent from the particular process at hand.
Any four-dimensional numerator function $N(q)$ can be rewritten by reconstructing $4$-dimensional denominators $\d{i} = ({q} + p_i)^2-m_i^2$ where $p_i$ are linear combinations of the incoming and outgoing four-momenta and $q$ is the integration momentum. The universal functional form of such decomposition is process-independent~\cite{Ossola:2006us}. 
After this algebraic operation, all terms in $N(q)$ that, aside from reconstructed denominators, still depend on $q$ have to vanish upon integration~\cite{delAguila:2004nf}  and therefore do not contribute to the scattering amplitude. Such terms are called ``spurious terms''~\cite{Ossola:2006us}. The physical content of the Feynman integral lies in the other terms, namely the part of the numerator decomposition in which $q$-independent functions of masses and four-momenta multiply sets of reconstructed denominators. In this framework, the computation of Feynman integrals is remapped into the purely algebraic problem of the extraction of such coefficients. The four-dimensional integrand-level reduction algorithm has been implemented in the code {\cuttools}~\cite{Ossola:2007ax}, that is publicly available.

The appearance of divergences in the evaluation of Feynman integrals requires the use of a regularization technique. In dimensional regularization, the integration momentum is upgraded to dimension $d = 4 - 2 \epsilon$. Such procedure is responsible for the appearance of the rational part. 

Following the OPP approach, there are two contributions to the rational term, which have different origins: the first contribution, called  ${\cal R}_1$, appears from the mismatch between the $d$-dimensional denominators of the scalar integrals and the $4$-dimensional denominators and can be automatically computed by means of a fictitious shift in the value of the masses~\cite{Ossola:2006us,Ossola:2007ax}. A second piece, called ${\cal R}_2$, comes directly from the $d$-dimensionality of the numerator function, and can be  recovered as tree-level calculations by means of \emph{ad hoc} model-dependent Feynman rules ~\cite{Ossola:2008xq,Draggiotis:2009yb}.

\paragraph{$D$-dimensional Integrand Reduction} 
Since the rational term escapes four-dimensional detection, significant improvements have been achieved performing the integrand decomposition directly in dimension $d = 4 - 2 \epsilon$\ rather than four~\cite{Ellis:2007br, Mastrolia:2010nb}, which indeed allows for the combined determination of all contributions at once~\cite{Giele:2008bc}.

These ideas led to the development of a new algorithm, called {\samurai}~\cite{Mastrolia:2010nb}, in which the polynomial structures described above also include a dependence on the extra-dimensional parameter $\mu$ needed for the automated computation of the full rational term according to the $d$-dimensional approach, the parametrization of the residue of the quintuple-cut in terms of the extra-dimensional scale \cite{Melnikov:2010iu} and the sampling of the multiple-cut solutions via Discrete Fourier Transform~\cite{Mastrolia:2008jb}.

The integrand-reduction algorithm was originally developed for renomalizable gauge theories at one-loop, namely Feynman integrals in which the the rank of the numerator function never exceeds the number of external legs. In order to deal with more general models, such as effective theories or certain BSM scenarios, this restriction should be lifted. In particular,  as preparatory work for the evaluation of $pp \to H +2,3$ jets in gluon fusion~\cite{vanDeurzen:2013rv, Cullen:2013saa}, where effective gluon vertices generated by the large top-mass limit appear and trigger higher rank terms,  {\samurai} has been enhanced to allow for such an extension~\cite{Mastrolia:2012bu, vanDeurzen:2013pja}.

\paragraph{Integrand Reduction via Laurent Expansion}  
If the analytic form of the numerator is known or, more generally, if
the polynomial dependence on the loop momentum is known, all
coefficients in the integrand decomposition can be extracted by
performing a Laurent expansion (with respect to one of the free
parameters which appear in the solutions of the cuts) implemented via
polynomial division~\cite{Mastrolia:2012bu}.
%If the analytic form of the numerator is known, all coefficients in the integrand decomposition can be extracted by 
%performing a Laurent expansion with respect to one of the free parameters which appear in the solutions of the cuts~\cite{Mastrolia:2012bu}. 
This idea provides a different and very powerful approach to integrand reduction, and allows for an efficient and precise evaluation of all the coefficients. Moreover, the contributions coming from the subtracted terms can be implemented as analytic corrections, replacing the numerical subtractions of the original algorithm. The parametric form of these corrections can be computed once and for all, in terms of a subset of the higher-point coefficients required by the original algorithm.

%If either the analytic expression of the integrand or the tensor structure of the numerator is known, the coefficients of the Laurent
%expansion can be computed by performing a polynomial division between the
%numerator and the set of denominators. 
The method has been implemented in the \C++ library {\Ninja}~\cite{Peraro:2014cba}. Its use within the {\Gosam} framework  showed an exceptional improvement  in  the computational performance~\cite{vanDeurzen:2013saa}, both in terms of speed and precision, with respect to the standard algorithms. The  {\Ninja}  library has been already employed in several calculation, among them the evaluation of NLO QCD corrections to $p p \to t {\bar t} H j $~\cite{vanDeurzen:2013xla}. It has also been interfaced within \Formcalc~\cite{Gross:2014ola} and very recently within \mgamc~\cite{Alwall:2014hca}.

\section{NLO Phenomenology with \gosamtwo}

The {\gosam} framework~\cite{Cullen:2011ac, Cullen:2014yla} combines automated diagram generation, algebraic manipulation~\cite{Nogueira:1991ex}, tensorial decomposition, and integrand reduction. 
After the automated generation of all Feynman integrals contributing to the selected process, the virtual corrections can be evaluated using the integrand reduction via Laurent expansion~\cite{Mastrolia:2012bu} provided by {\Ninja}, which is the default choice, or the $d$-dimensional integrand-level reduction method, as implemented in \SAMURAI~\cite{Mastrolia:2010nb}, or alternatively the tensorial decomposition provided by {\Golem}~\cite{Binoth:2008uq,Heinrich:2010ax}. The only task required from the user is the preparation of an input file for the generation of the code and the selection of the various options, without having to worry about  the internal details. \gosamtwo, which was released in 2014, is a new version of the code that offers numerous improvements on both generation and reduction, resulting in faster and  more stable codes for calculations within and beyond the Standard Model. 
 {\gosamtwo} also contains the extended version of the standardized \emph{Binoth Les Houches Accord} (BLHA) interface~\cite{Binoth:2010xt,Alioli:2013nda} to Monte Carlo programs.

The computation of physical observables at NLO accuracy, such as cross sections and differential distributions, requires to combine the one-loop results for the virtual amplitudes obtained with \gosam{}, with other tools that can take care of the computation of the real emission contributions and of the subtraction terms, needed to control the cancellation of IR singularities. This can be obtained by embedding the calculation of virtual corrections within a Monte Carlo framework (MC), that can also provide the phase-space integration, and the combination of the different pieces of the calculation. A complete table of \gosam's interfaces with MC programs has been recently presented in~\cite{vanDeurzen:2014uaa}.

While in the following we will describe two Higgs-related projects, developed within the frameworks of \Sherpa~\cite{Gleisberg:2008ta} and  \amcnlo\ respectively, other significant NLO studies have been performed with \gosam\  during the past year, related to di-Higgs production in association with jets~\cite{Dolan:2015zja}, electroweak corrections to the production of a vector boson plus jets~\cite{Chiesa:2015mya}, and $W b \bar{b} j$ production at hadron colliders~\cite{Luisoni:2015mpa} implemented in the \PB\ framework~\cite{Alioli:2010xd}. 

\paragraph{Higgs Boson Production in Gluon Fusion}  

As a first example of application of \gosam\, we will summarize the efforts that allowed us to complete the challenging calculation of NLO QCD corrections to the associated production of a Higgs boson and three jets at the LHC in gluon fusion in the large top-mass limit~\cite{Cullen:2013saa}.

In this approximation, the coupling of the Higgs boson to gluons, which in the full theory is mediated by a top-quark loop, is described by an effective operator that gives rise to vertices involving the Higgs field and up to four gluons. The presence of these new vertices leads to Feynman integrals in which the rank of the integration momentum in the numerator functions exceeds the number of denominators. As a consequence, all reduction algorithms needed to be upgraded in order to deal with such higher-rank integrals~\cite{Mastrolia:2012bu, vanDeurzen:2013pja, Guillet:2013msa}. The upgraded algorithms were tested by computing  $pp \to H +2$ jets in gluon fusion~\cite{vanDeurzen:2013rv}.

The complexity and the huge number of diagrams contained in $pp\to H+3$ jets, in which the one-loop virtual part alone involves more than ten thousand Feynman diagrams with up to rank-seven hexagons, required the {\gosam} code to be further enhanced. The introduction of numerical polarization vectors and the option to sum diagrams sharing the same propagators algebraically during the generation of the code led to an enormous gain in generation time and reduction of code size. Moreover, the optimized algebraic manipulation provided by {\sc Form 4.0}~\cite{Kuipers:2012rf} further helped to improve the performance. 
Concerning the reduction, a more stable and efficient extraction of all coefficients was achieved thanks to the use of the already mentioned \Ninja\ algorithm.

This calculation was also challenging for what concerns the real-emission contributions and the integration over phase space.
Due to the complexity of the integration, the first published results~\cite{Cullen:2013saa} were obtained with a hybrid setup which combined \GOSAM{} with both \Sherpa{} and the an \emph{in house} implementation of the MadDipole/Madgraph4/MadEvent framework~\cite{Frederix:2008hu}.  

An updated analysis obtained interfacing \GOSAM{} with \Sherpa, which contains results and distributions based on a set of ATLAS-like cuts and a comparison with the NLO predictions for $H+2$ jets, was published in the ``Physics at TeV Colliders: Standard Model Working Group Report''~\cite{Butterworth:2014efa}. More recently, new phenomenological analyses have been presented~\cite{Greiner:2015jha} which include numerical results for a large variety of observables for both standard cuts and VBF selection cuts. For more details, we refer the reader to the the talk of Nicolas Greiner at this conference~\cite{nicoloop15}. 

\paragraph{NLO QCD corrections to $pp \to$ \ttyy{}} 
%\paragraph{Interfacing \GOSAM{} and \mgamc} 
As a second application, we present a new interface that was developed between the multipurpose Monte Carlo tool \mgamc\ and \gosam~\cite{vanDeurzen:2015cga}.  
On the one hand, this tandem allows the user of \amc{} to switch
between two options, namely between the default code {\sc
MadLoop} \cite{Hirschi:2011pa} fully integrated directly in the MC
distribution package, and \Gosam{}. Thus, the user can experience the
evaluation of NLO virtual corrections by means of two alternative
solutions corresponding to different algorithms and methods of
generation and evaluation of Feynman amplitudes.  On the other
hand, \Gosam{} is interfaced to other Monte Carlo codes beside \amcnlo{}, therefore the user of the MCs can explore and compare the different features of the event generators, without being biased by the performances of the one-loop provider, since they all can be run using \Gosam{}.
%The interface between \Gosam{} and \amcnlo{} is based on the
%standards of the first BLHA~\cite{Binoth:2010xt}.
%When running the \mgamc{} interactive session, the command ``\texttt{set OLP GoSam}''
%changes the employed OLP from its default \MadLoop{} to \gosam{}. 
%Alternatively, the file \texttt{input/mg5\_configuration.txt}
%can be edited to include the line \texttt{OLP = GoSam}.\\
%The interface is available starting
%from \amcnlo{} version 2.3.2.2. 

The interface between \Gosam{} and \amcnlo{} is based on the
standards of the first BLHA~\cite{Binoth:2010xt}.
When running the \mgamc{} interactive session, the command ``\texttt{set OLP GoSam}''
changes the employed OLP from its default \MadLoop{} to \gosam{}.
Alternatively, the user should include the line \texttt{OLP = GoSam} by editing the file \texttt{input/mg5\_configuration.txt}.
The interface is available starting
from \amcnlo{} version 2.3.2.2. 

\begin{figure}[ht] 
\bcen
%\subfigure[]
{\includegraphics[width=7.5cm]{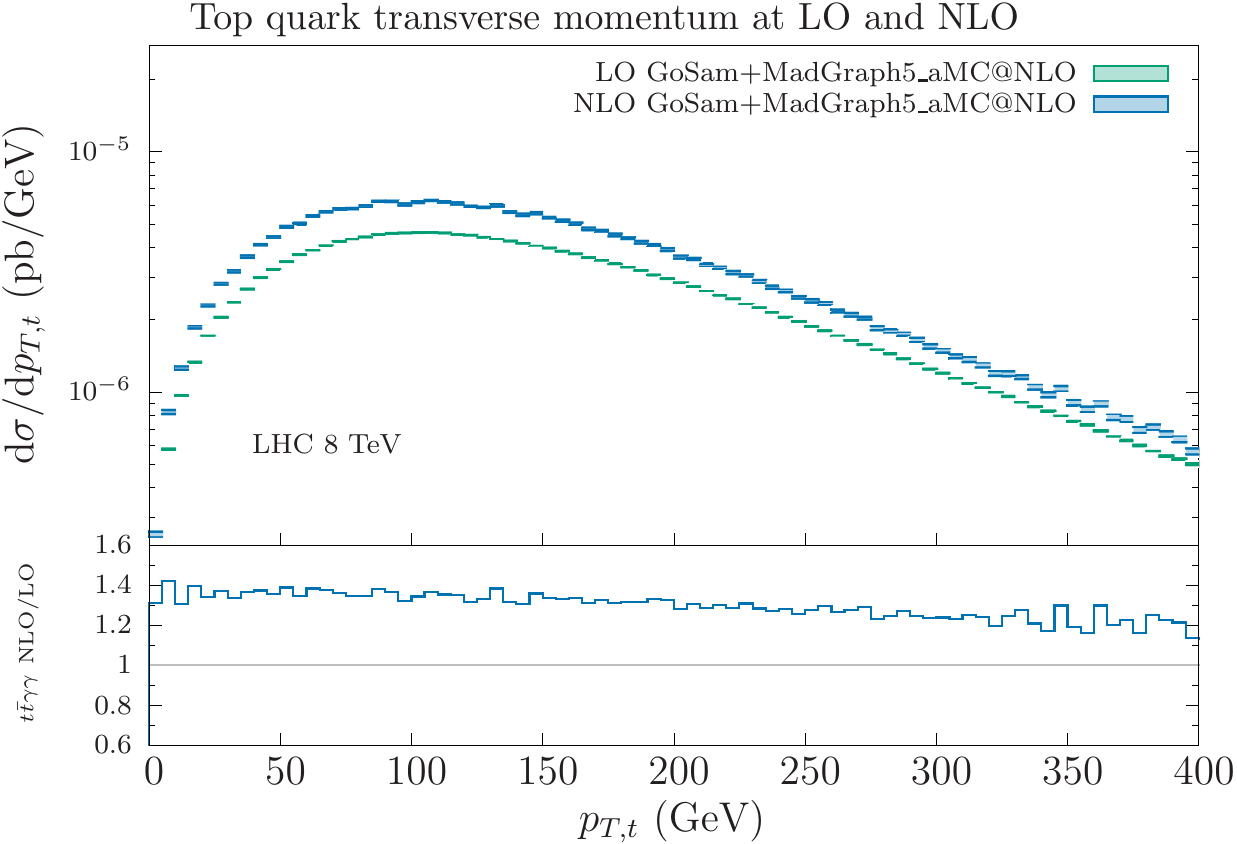}} 
\hspace{0.31cm}
 {\includegraphics[width=7.5cm]{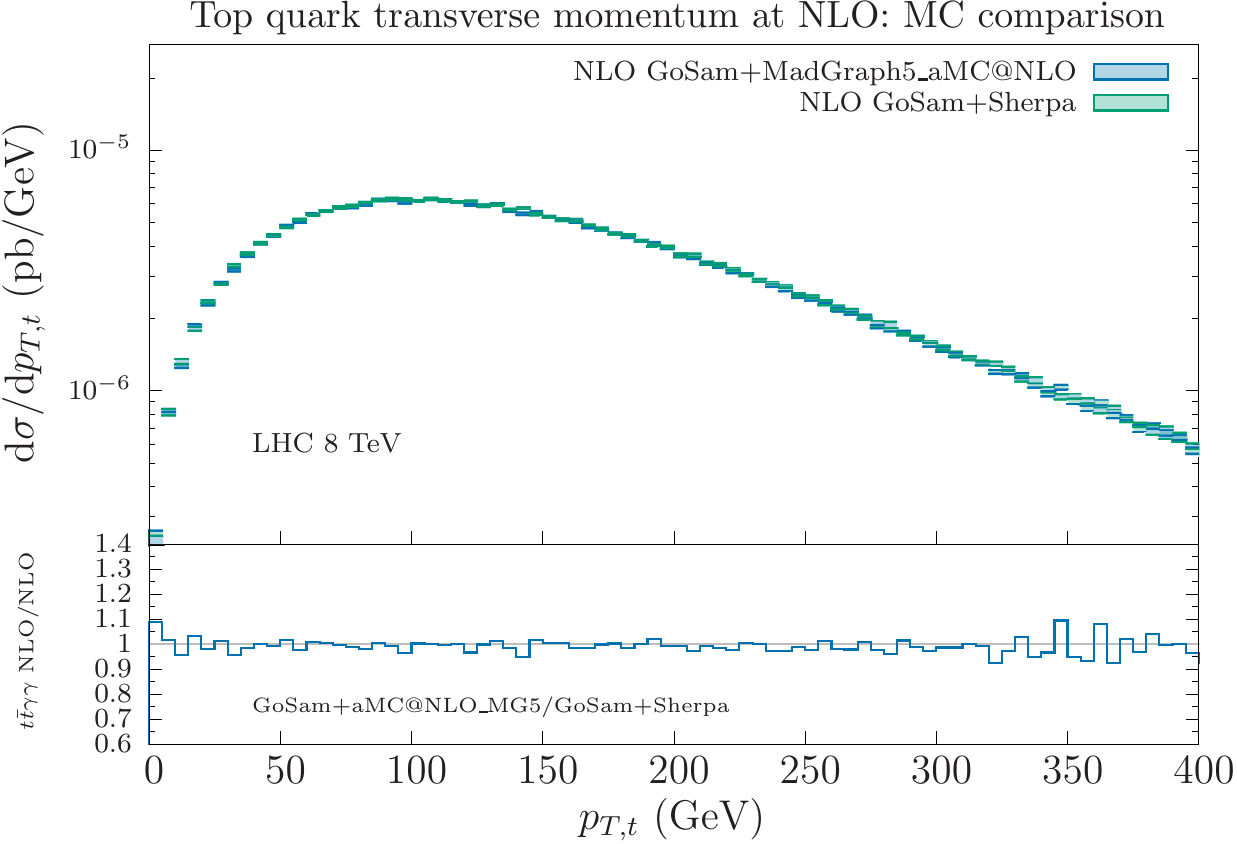}}\\
\ecen
\caption{Transverse momentum of the top quark in $ p p \to t \bar t \gamma \gamma$  for the LHC at 8 TeV: 
LO and NLO distributions for the transverse momentum of the top quark (left) and NLO comparison between \GOSAM{}+\amcnlo\ and \GOSAM{}+\Sherpa\ (right).  }
\label{fig:ttyy}
\end{figure}

To validate the interface several cross checks were performed.  The
loop amplitudes of \gosam{} and \madloop{} were compared for
single phase space points and also at the level of the total cross
section for a number of different processes, as presented in a
dedicated table in~\cite{hansthesis}. Furthermore, for $pp\rightarrow
t\bar{t}\gamma\gamma$, a fully independent check was also performed by
computing the same distributions using \gosam{} interfaced
to \Sherpa{} (see Figure~\ref{fig:ttyy}). %As an illustration, in Figure~\ref{fig:ttyy}, we report  distributions for the transverse momentum of the top quark. 
%obtained with and the comparison with \GOSAM{}+\Sherpa.  %All predictions are computed for a center of mass energy of 8 TeV.

As an application of this novel framework, we computed the NLO QCD corrections to $pp \to$ \ttH{} and $pp \to$ \ttyy{} matched to a parton shower~\cite{vanDeurzen:2015cga}. The study is performed using NLO predictions for \ttH\ and continuum \ttyy{} production. The top and anti-top quarks are subsequently decayed semi-leptonically  with \Madspin~\cite{Frixione:2007zp}, taking into
account spin correlation effects, and then showered and hadronised by means of  \Pythia~8.2~\cite{Sjostrand:2014zea}.
We compared several distributions to disentangle the two processes and focused in particular on
observables designed to study spin correlation effects. While NLO corrections are sizable and provide a clear reduction of theoretical uncertainties, they only mildly distort the shape of the various distributions.

\section{\gosam\ beyond NLO}

As the focus of the theoretical particle community is shifting towards NNLO calculations, automated codes for the generation and reduction of two-loop virtual amplitudes, as well as for the computation of real-virtual contributions, will be much needed. It is not yet clear whether at NNLO the most efficient way to proceed will be targeting full process-independent automation, as achieved at NLO, or rather developing ad hoc tools for specific classes of processes. While \gosam\ was initially designed and developed to compute one-loop virtual contributions needed by NLO predictions, several of it features can be adapted and extended to address specific tasks needed by higher order calculations. 

Concerning the generation virtual two-loop matrix elements, the routines in \gosam\  have been extended to produce the full list and expressions for all two-loop Feynman diagrams contributing to any process: as for the one-loop case, the code depicts all contributing diagrams as output on file, takes care of the algebra by means of \form, and projects the expressions over the appropriate tensor structures, to extract the form factors. Interfaces to codes for the reduction to master integrals are currently in progress.
%{\sc Reduze}~\cite{}, LiteRed~\cite{}, FIRE~\cite{} are in progress.

\gosamtwo\ has been already used within NNLO calculations, for the evaluation of the real-virtual contributions. In particular, the code has been employed in the evaluation of electroweak production of top-quark pairs in electron-positron annihilation at NNLO in QCD~\cite{Gao:2014nva, Gao:2014eea}, and also to cross-check numerically the analytic expressions obtained within the evaluation of Higgs boson decay into b-quarks at NNLO accuracy~\cite{DelDuca:2015zqa}.

As a last application beyond NLO, \gosam\ has been used for the evaluation of associated production of a top-quark pair and a Higgs boson at approximate NNLO in QCD~\cite{Broggio:2015lya}.
In this paper,  approximate formulas were obtained by studying soft-gluon corrections in the limit
 where the partonic center-of-mass energy approaches the invariant mass
 of the $t\bar{t}H$ final state, where the latter can be arbitrarily  large.  
 The approximate NNLO corrections are extracted from the
 perturbative information contained in a soft-gluon resummation formula valid to NNLL accuracy, whose
 derivation is based on SCET (for a recent review, see~\cite{Becher:2014oda}). 
 The soft-gluon resummation formula for this process contains three essential ingredients, all of which are matrices in the
 color space needed to describe four-parton scattering: a hard
 function, related to virtual corrections; a soft function, related
 to real emission corrections in the soft limit; and a soft
 anomalous dimension, which governs the structure of the all-order
 soft-gluon corrections through the renormalization group (RG).
Of these three ingredients, both the NLO soft function
 \cite{Ahrens:2010zv} and NLO soft anomalous dimension
 \cite{Ferroglia:2009ep} needed for NNLL resummation
 in processes involving two massless and two massive partons  can be adapted directly to $t\bar{t}H$ production.
The NLO hard function is instead process-dependent and was evaluated by using modified versions of the \Gosam, \MadLoop, and \openloops~\cite{Cascioli:2011va} codes.

\section{Integrand Reduction beyond One Loop}

The idea of applying the integrand reduction to Feynman integrals beyond one-loop, first applied in Refs.~\cite{Mastrolia:2011pr, Badger:2012dp}, has been the target of several studies in the past five years, thus providing a new promising direction in the study of multi-loop amplitudes. 
%~\cite{, Zhang:2012ce, Mastrolia:2012an,Kleiss:2012yv, Badger:2012dv,Feng:2012bm, Mastrolia:2012wf, Mastrolia:2013kca, Feng:2014nwa, Ita:2015tya, Badger:2015lda}, thus 

As a basic ingredient for the integrand reduction, a proper parametrization of the residues at the multi-particle poles of higher-loop integrals in needed~\cite{Mastrolia:2011pr}. Like the one-loop case, the parametric form of the polynomial residues should be process-independent and determined once for all from the corresponding multiple cut. Unlike the one-loop case however, the basis of master integrals beyond one-loop is not straightforward. Moreover, the splitting between ``spurious'' and ``physical'' terms in the residues is more tricky due to the presence irreducible scalar products, namely scalar products involving integration momenta that cannot be reconstructed in terms of denominators. 

In Refs.~\cite{Zhang:2012ce, Mastrolia:2012an}, the determination of the residues at the multiple cuts has been systematized as a problem of multivariate polynomial division in algebraic geometry. The use of these techniques allowed to apply the integrand decomposition not only at one loop, as originally formulated, but at any order in perturbation theory. 
Moreover, this approach confirms that the shape of the residues is uniquely determined by the on-shell conditions, without any additional constraint. 

The algorithm presented in Ref.~\cite{Mastrolia:2012an} allows to decompose a general loop integrand by means of a  powerful recurrence relation. In general, if the on-shell conditions have no solutions, the integrand is {\it reducible}, namely it  can be written in terms of lower point functions. One example of this class of integrands are the six-point functions at one loop, which are fully reducible to lower point functions, as well known for a long time~\cite{Kallen:1964zz}. When the on-shell conditions admit solutions, the corresponding residue is obtained dividing the numerator function modulo the Gr\"obner basis of the corresponding cut.
The {\it remainder} of the division provides  the {\it residue}, while the {\it quotients} generate integrands with less denominators. 

\begin{figure}[h!]
\begin{center}
\includegraphics[width=0.67\textwidth]{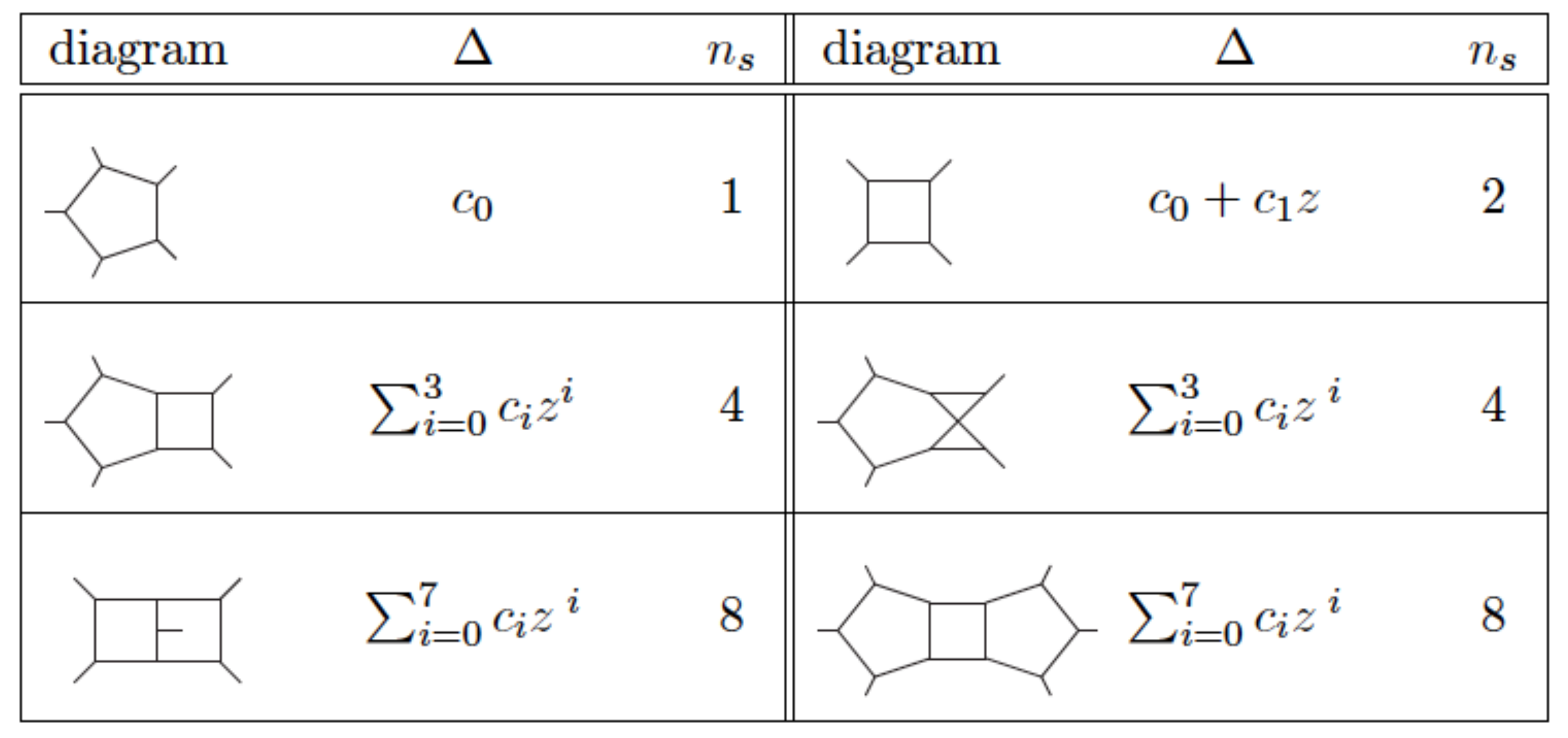}
\end{center}
\caption{
Examples of maximum cuts. With the exception of the first diagram in the left column, which represents the 5-ple cut of the 5-point 
one-loop dimensionally regulated amplitude, all the other diagrams in the table are considered 
in four dimensions.}
\label{maxcut}
\end{figure}

A fundamental ingredient that guarantees the consistency of the integrand reduction approach beyond one-loop is contained in the {\it Maximum Cut Theorem}~\cite{Mastrolia:2012an}. After defining {\it Maximum-cuts} as the maximum number of on-shell conditions which can be simultaneously satisfied by the loop momenta, the {\it Maximum Cut Theorem} ensures that the corresponding residues can always be reconstructed by evaluating the numerator at the solutions of the cut, since they are parametrized by exactly $n_s$ coefficients, where $n_s$ is the number of solutions of the multiple cut-conditions.  This theorem extends at all orders the features of the one-loop quadruple-cut in dimension four~\cite{Britto:2004nc,Ossola:2006us}, in which two complex solutions of the cut allow for the determination of the two coefficients needed to parametrize the residue. 

In Figure~\ref{maxcut}, we show the structures of the residues of the maximum cut, together with the corresponding 
values of $n_s$, for a selection of diagrams with different number of loops.  
For each case, the general structure of the residue $\Delta$ and the corresponding  value of $n_s$ are provided~\cite{Mastrolia:2012an}.  Similar conditions can be found for more complicated topologies at higher loops. 

The integrand recurrence relation described above can be applied in two ways~\cite{Mastrolia:2013kca}. Given the form of all residues, the coefficients which appear in the residues can be determined by evaluating the numerator at the solutions of the multiple cuts, as many times as the number of the unknown coefficients. This approach, which in the paper we labeled \emph{fit-on-the-cuts}, has been employed at one loop in the original integrand reduction~\cite{Ossola:2006us}, and the language of multivariate polynomial division provides its generalization at all loops. 

As a very different strategy~\cite{Mastrolia:2013kca}, which we dubbed \emph{divide-and-conquer}, the decomposition can be obtained analytically by means of polynomial divisions. This approach does not require prior knowledge of the parametric form of the residues or the solutions of the multiple cuts, and the reduction algorithm is applied directly to the expressions of the numerator functions. It is worth noticing that this strategy can be successfully applied to integrands with denominators appearing with multiple powers, which represented a long-standing problem within unitarity-based methods. 

\section{Conclusions and Future Outlook}

Integrand-reduction techniques played an important role in the development of automated codes for NLO calculations. Algorithms such as the integrand-level OPP reduction, $D$-dimensional integrand reduction, integrand reduction via Laurent expansion, which we described in the first part of this presentation, are now embedded and interfaced within Monte Carlo tools that allow users to compute cross sections and distributions for a wide variety of processes at NLO accuracy, as needed by the LHC experimental collaborations.

The developments of the past decade also showed how a better understanding of the mathematical properties of scattering amplitudes eventually provides the foundations for the construction of efficient algorithms for their evaluation.  Looking ahead, as the focus of the community is shifting towards the challenges presented by NNLO calculations, new ideas and techniques, along with improved version of known algorithms, will represent alternatives, cross-checks, and ultimately new solutions to known problems. In this context, it will be interesting to observe whether the extensions of integrand-level techniques to higher orders will succeed to provide a comparable level of reliability, and eventually of automation, as in the one-loop case, and to what extent the \gosam{} framework could be extended to explore the new frontiers in precision calculations.

\paragraph{Acknowledgments} I would like to thank the present and former members of the \gosam\ Collaboration for their many contributions to the results presented in this talk. I would also like to acknowledge Alessandro Broggio, Andrea Ferroglia, Rikkert Frederix, and Valentin Hirschi for stimulating discussions and for providing new opportunities for the use of the \gosam\ framework. \\ Work supported in part by the National Science Foundation under Grants  PHY-1068550 and PHY-1417354 and by the PSC-CUNY Awards No. 67536-00 45 and No. 68687-00 46. 

%\bibliographystyle{JHEP}
%\bibliography{references}

\providecommand{\href}[2]{#2}\begingroup\raggedright\endgroup

\end{document}